\documentclass[conference]{IEEEtran}
\ifCLASSINFOpdf
\else
\fi
\usepackage{amssymb}
\usepackage{indentfirst}
\usepackage{amsmath}
\usepackage{bm}
\usepackage{booktabs}
\usepackage{graphicx}
\usepackage{subfigure}
\usepackage{epsfig}
\usepackage{algorithm}
\usepackage[noend]{algorithmic}  
\usepackage{cite}
\usepackage{multirow}

\usepackage{changepage}
\begin{document}
\title{Segmentation-Discarding Ordered-Statistic Decoding for Linear Block Codes}

%\author{\IEEEauthorblockN{Michael Shell}
%\IEEEauthorblockA{School of Electrical and\\Computer Engineering\\
%Georgia Institute of Technology\\
%Atlanta, Georgia 30332--0250\\
%Email: http://www.michaelshell.org/contact.html}
%\and
%\IEEEauthorblockN{Homer Simpson}
%\IEEEauthorblockA{Twentieth Century Fox\\
%\and
%\IEEEauthorblockN{James Kirk\\ and Montgomery Scott}
%\IEEEauthorblockA{Starfleet Academy\\
%San Francisco, California 96678--2391\\


\author{\IEEEauthorblockN{Chentao Yue,
Mahyar Shirvanimoghaddam,
Yonghui Li, Branka Vucetic}
\IEEEauthorblockA{\textit{School of Electrical and Information Engineering}\\
\textit{The University of Sydney},
NSW, Australia\\
Email: \{chentao.yue, mahyar.shirvanimoghaddam, yonghui.li, branka.vucetic\}@sydney.edu.au}
}


\maketitle

\begin{abstract}
In this paper, we propose an efficient reliability based segmentation-discarding decoding (SDD) algorithm for short block-length codes. A novel segmentation-discarding technique is proposed along with the stopping rule to significantly reduce the decoding complexity without a significant performance degradation compared to ordered statistics decoding (OSD). In the proposed decoder, the list of test error patterns (TEPs) is divided into several segments according to carefully selected boundaries and every segment is checked separately during the reprocessing stage. Decoding is performed under the constraint of the discarding rule and stopping rule. Simulations results for different codes show that our proposed algorithm can significantly reduce the decoding complexity compared to the existing OSD algorithms in literature.
\end{abstract}


% \fi
\IEEEpeerreviewmaketitle

\section{Introduction}
Since 1948, when Shannon introduced the notion of channel capacity and channel coding\cite{Shannon}, researchers have been looking for powerful channel codes which can approach the Shannon capacity. Low density parity check (LDPC) and turbo codes have been demonstrated to be capacity approaching and been widely applied in 3G (3rd-generation) and 4G (4th-generation) mobile communications\cite{lin2004ECC}. The polar code proposed by Erdal Arikan et al. in 2008 \cite{polar2009} also attracted much attention in the last decade and has been chosen as one of the standard coding schemes for 5G (5th-generation) communications. In addition to large bandwidth and high-speed enhanced Mobile BroadBand (eMBB) scenarios, 5G has put forward the demand for ultra-reliable and low latency communications (uRLLC). For uRLLC, reducing the latency mandates the use of short block-length codes and conventionally designed moderate/long codes are not suitable \cite{shirvanimoghaddam2018short}. Short code design and the related decoding algorithms have rekindled a great deal of interest among industry and academia recently\cite{liva2016codeSurvey,2016Performancecomparison}.

Ordered statistics decoding (OSD) was proposed in 1995 as an approximate maximum likelihood (ML) decoder for block codes\cite{Fossorier1995OSD}. OSD has recently aroused interests again because its potential to be a universal decoding algorithm for all short block-length codes. For a linear block code $(N,K)$ with minimum distance $d_{min}$, it is proven that an OSD with order $m = \lceil d_{min}/4-1\rceil$ is asymptotically optimum \cite{Fossorier1995OSD}. However, the decoding complexity is a main disadvantage of OSD, as an order-$m$ OSD needs a candidate list with size of $\sum_{l=1}^{m}\binom{K}{l}$ and the overall algorithmic complexity can be up to $O(K^m)$. 

Much previous work has focused on improving OSD in terms of efficiency and some remarkable progresses have been achieved\cite{Fossorier1996OSDimprovement,Fossorier2002IISR,FossorierBoxandMatch,Fossorier2007OSDbias,wu2007preprocessing_and_diversification,Wu2007OSDMRB,jin2006probabilisticConditions,2012Segment}. The Box-and-Match algorithm \cite{FossorierBoxandMatch} can greatly reduce the size of the candidates list, while it brings other computations due to the matching process. Decoding with different biases over reliability was proposed in \cite{Fossorier2007OSDbias} to refine the performance, and skipping and stopping rules were used in \cite{wu2007preprocessing_and_diversification} and \cite{Wu2007OSDMRB} to abandon unpromising candidates. All of the above methods can be combined with the iterative information set reduction (IISR) technique in \cite{Fossorier2002IISR} to further reduce the complexity. Recently an approach proposed in \cite{2012Segment} cuts the most reliable basis (MRB) to several partitions and performs independent OSD over each of them, but it overlooks candidates generated across partitions so that a dramatic performance degradation is resulted. Also a fast OSD algorithm combining stopping rules from \cite{Wu2007OSDMRB} and sufficient conditions from \cite{jin2006probabilisticConditions} was proposed in \cite{NewOSD-5GNR}, which can reduce the complexity from $O(K^m)$ to $O(K^{m-2})$ in high signal-to-noise ratios (SNRs).

In this paper, we propose a new fast decoding algorithm combining segmentation-discarding technique and an easily calculated stopping rule. Firstly the list of test error patterns (TEPs) is partitioned into $Q$ segments according to $Q+1$ carefully selected boundaries over MRB. Then a segment in each reprocessing is discarded if it satisfies a discarding rule. The rule estimates the reliability of each segment by calculating a lower bound on the distance from received signal to the decoded codeword. Reprocessing is performed from the segment with highest priority and terminated if all segments are checked or a stopping rule is satisfied. This algorithm can achieve the performance of the OSD algorithm of large orders with significantly reduced complexity. Simulation results show that this degree of complexity reduction maintains for any rate eBCH code. In addition, the complexity and memory overhead due to the segmentation and discarding is negligible.

The rest of this paper is organized as follows: Section \ref{Preliminaries} describes preliminaries. In Section \ref{MainPart}, the proposed segmentation-discarding algorithm is presented. The analysis of computation complexity is provided in Section \ref{Computational Complexity}. Simulation results are  presented in Section \ref{Simulation} and conclusions are drawn in Section \ref{Conclusion}.

\section{Preliminaries} \label{Preliminaries}
We consider a binary linear block code $C(N,K)$, where $K$ and $N$ denote the information block size and codeword length, respectively. Let $ {\bf b} = [ b_{1},b_{2}\ldots,b_{K} ] $ and ${\bf c} = [c_{1},c_{2}\ldots,c_{N}] $ denote the information sequence and codeword, respectively. Given the generator matrix ${\bf G}$, the encoding  operation can be described as ${\bf c} = {\bf b}\cdot{\bf G}$. 

We suppose an additive white Gaussian Noise (AWGN) channel and binary phase shift keying (BPSK) modulation. Let $\mathbf{s} = [s_{1},s_{2},\ldots,s_{N}]$ denote the modulated signals, where $s_{i} = (-1)^{c_{i}}\in \{\pm 1\}$. At the channel output, the received signal is given by $\mathbf{r} = \mathbf{s} + \mathbf{n}$, where $\mathbf{n}$ is the vector of white Gaussian noise samples with zero mean and variance $N_{0}/2$.

In general, if the codewords in $C$ have equal transmission probability, the log-likelihood-ratio of the $i$-th symbol of the received signal can be calculated as $\delta_{i} 
\triangleq \ln \frac{P_{r}(c_{i}=1|r_{i})}{P_{r}(c_{i}=0|r_{i})}$, which can be further simplified to $\delta_{i} = 4r_{i}/N_{0}$ \cite{Wu2007OSDMRB}. Bitwise hard decision can be used to obtain the codewords estimation $\mathbf{y}=[y_{1},y_{2},\ldots,y_{N}]$ according to following rule:
\begin{equation} \label{HD rule}
y_{i}=
\begin{cases}
1& \text{for} \ r_{i}<0\\
0& \text{for} \ r_{i}\geq 0
\end{cases}
\end{equation}
where $y_{i}$ is the estimation of codeword bit $c_{i}$.

We consider the scaled magnitude of log-likelihood-ratio as the reliability (or confidence value) corresponding to bitwise decision, defined as $\alpha_{i} = |r_{i}|$. Utilizing the bits reliability, the soft-decision decoding can be effectively conducted using the OSD algorithm \cite{Fossorier1995OSD}. At the first step of OSD, a permutation $\pi_1$ is performed to sort the received signals $\mathbf{r}$ and the corresponding columns of generator matrix in descending order of their reliabilities. Thus the sorted received signal vector is $\mathbf{r}'=\pi_{1}(\mathbf{r})$ and the corresponding reliability vector $\bm{\alpha}' = [\alpha_{1}',\alpha_{2}',\ldots,\alpha_{N}'] $ satisfies
\begin{equation}
\alpha_{1}' \geq \alpha_{2}' \geq \ldots \geq \alpha_{N-1}' \geq \alpha_{N}' \text{.}
\end{equation}

 Next, the systematic form matrix $\mathbf{\widetilde G} = [\mathbf{I}_{K} \  \mathbf{\widetilde{P}}]$ is obtained by performing Gaussian elimination on $\mathbf{G}' = \pi_1(\mathbf{G})$, where $\mathbf{I}_{K}$ denotes the K-dimensional identity matrix and $\mathbf{\widetilde{P}}$ is the parity sub-matrix. An additional permutation $\pi_{2}$ may be necessary during Gaussian elimination to ensure that the first K columns are linearly independent. Correspondingly, the received signal and reliability are finally sorted to $\mathbf{\widetilde r} = \pi_{2}(\pi_{1}(\mathbf{r}))$ and $\bm{\widetilde \alpha} = \pi_{2}(\pi_{1}(\bm{\alpha}))$, respectively. A simple greedy search algorithm to perform the permutation $\pi_{2}$ can be found in \cite{Fossorier1995OSD}. 

After the transformation, the first $K$ index positions $[1,2,\ldots,K]$ are associated with the most reliable basis (MRB) \cite{Fossorier1995OSD}, and the rest of positions  $[K+1,K+2,\ldots,N]$ are associated with the redundancy part. For the phase-0 reprocessing, the hard decision is performed on ordered sequence $\mathbf{\widetilde r}$ using decision rule (\ref{HD rule}) to obtain estimated information $\mathbf{\widetilde y}$. Let $\mathbf{\widetilde y}_{B}$ denotes first $K$ positions of $\mathbf{\widetilde y}$ corresponding to MRB, so the first candidate codeword is obtained by re-encoding as
\begin{equation}
\mathbf{\widetilde c}_{0} = \mathbf{\widetilde y}_{B}\mathbf{\widetilde G}\text{.}
\end{equation}
Obviously, $\mathbf{\widetilde c}_{0}$ is the transmitted codewords if and only if there are no errors in MRB positions, otherwise, a test error pattern (TEP) $\mathbf{e} = [e_{1},e_{2},\ldots,e_{K}]$ is added to MRB hard-decision $\mathbf{\widetilde y}_{B}$ before re-encoding, which is equivalent to flipping bits corresponding to nonzero positions of $\mathbf{e}$. The decoding operation with bit flipping can be described as
\begin{equation}
\mathbf{\widetilde c}_{e} = \left(\mathbf{\widetilde y}_{B}\oplus \mathbf{e}\right)\mathbf{\widetilde G} = \left[\mathbf{\widetilde y}_{B}\oplus \mathbf{e} \  \left(\mathbf{\widetilde y}_{B}\oplus \mathbf{e}\right)\mathbf{\widetilde{P}}\right] \text{,}
\end{equation}
where $\mathbf{\widetilde c}_{e} = [{\widetilde c}_{e,1},{\widetilde c}_{e,2},\ldots,{\widetilde c}_{e,N}]$ is the candidates codewords with respect to TEP $\mathbf{e}$. 

In the reprocessing of OSD, a number of TEPs are checked to generate codeword candidates until a predetermined maximum candidate number is achieved. For BPSK modulation, finding the best ordered codeword estimation $\mathbf{\widetilde c}_{opt}$ is equivalent to minimizing the weighted Hamming distance (WHD) \cite{valembois2002comparison}, which is defined as
\begin{equation} \label{WHD}
\mathcal{D}_{e} \triangleq \sum_{\substack{0<i<N \\ \widetilde{c}_{e,i}\neq \widetilde{y}_{i}}} \widetilde{\alpha}_{i}\text{.}
\end{equation}
Finally, the optimal estimation $\hat{\mathbf{c}}_{opt}$ corresponding to initial received sequence $\mathbf{r}$ is obtained by performing inverse permutations over $\mathbf{\widetilde c}_{opt}$, i.e.
$\hat{\mathbf{c}}_{opt} = \pi_{1}^{-1}(\pi_{2}^{-1}(\mathbf{\widetilde c}_{opt}))$.

\section{A Fast OSD-based Decoding Algorithm} \label{MainPart}
 In this section, we propose a segmentation-discarding decoding (SDD) algorithm that can significantly reduce the decoding complexity, in which the TEP list is divided into several segments and some least reliable segments are discarded according to a discarding rule. In addition, the algorithm is terminated when a stopping condition is satisfied.
 
\subsection{Segmentation}

Firstly, the segmentation is conducted to prepare for applying the discarding rule. Specifically, all TEPs generated in phase-$l$ ($1\leq l \leq m$) reprocessing, denoted by $S_{l}$, is divided into $Q$ segments $\{S_{l_1},S_{l_{2}}\ldots S_{l_{Q}}\}$ according to $Q + 1$ boundaries $\{\beta_{0},\beta_{1},\beta_{2}\ldots,\beta_{Q}\}$ over MRB positions. The boundary position index satisfies
\begin{equation} \label{boundary position}
K+1 = \beta_{0} > \beta_{1} > \beta_{2} > \ldots > \beta_{Q-1}> \beta_{Q} = 1 \text{,}
\end{equation}
and the corresponding ordered reliabilities satisfies
\begin{equation} \label{BoundReliability}
{\widetilde \alpha}_{\beta_{0}}\leq {\widetilde \alpha}_{\beta_{1}}\leq {\widetilde \alpha}_{\beta_{2}}\leq \ldots \leq  {\widetilde \alpha}_{\beta_{Q-1}} \leq  {\widetilde \alpha}_{\beta_Q} \text{,}
\end{equation}
where ${\widetilde \alpha}_{\beta_{i}}$  is the ordered reliability of position $\beta_{i}$. The $i$-th segments bounded by $\beta_{i}$ is derived by
\begin{equation}  \label{SegmentDefine}
S_{l_i} = \left\{\mathbf{e} \in \{0,1\}^K \Big| w(\mathbf{e})=l, \, \sum_{j=\beta_i}^{\beta_0-1} e_{j} = l,\, \sum_{j=\beta_i}^{\beta_{i-1}-1} e_{j} \geq 1 \   \right\} \text{,}
\end{equation}
where $w(\mathbf{e})$ is the weight of TEP $\mathbf{e}$. The conditions in (\ref{SegmentDefine}) means that TEPs in the $i$-th segment $S_{l_{i}}$ only have nonzero elements over the positions from $\beta_{i}$ to $\beta_{0}$, and have at least one nonzero element over the positions from $\beta_{i}$ to $\beta_{i-1}$.

\begin{figure}
	\begin{center}
		\includegraphics[scale=0.16] {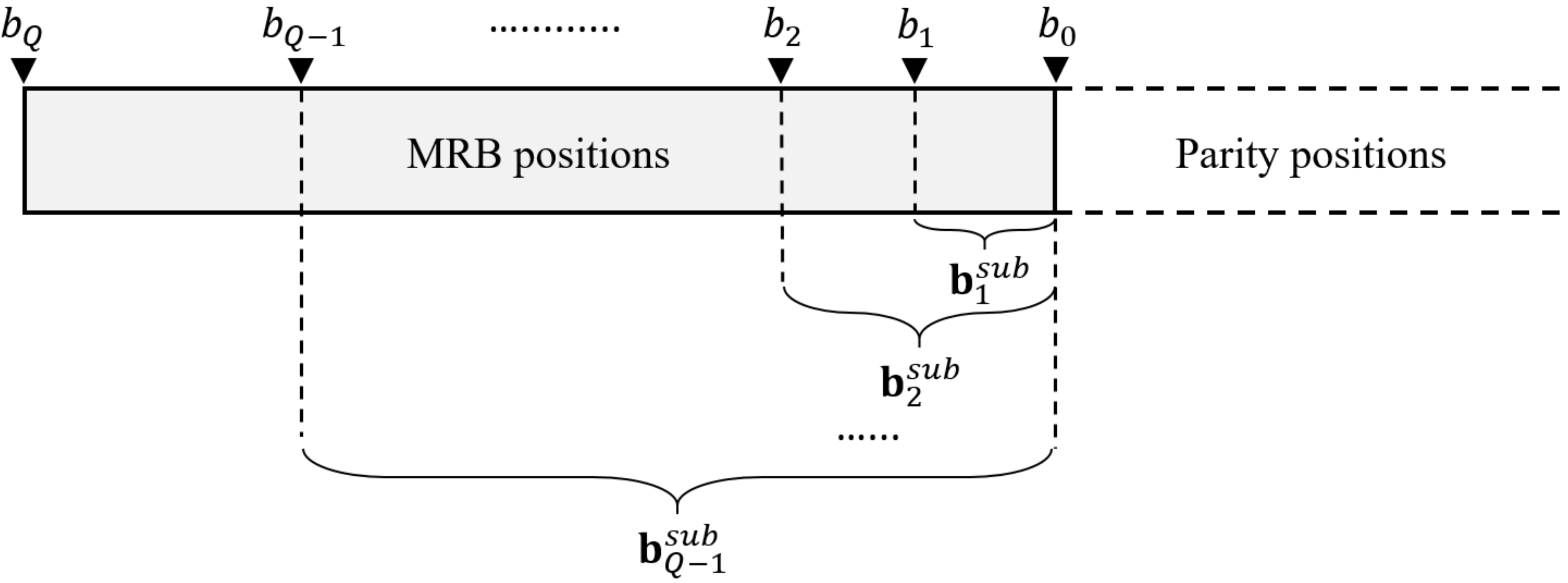}
		\caption{Segmentation of MRB positions}
		\label{MRBsegmentation}
	\end{center}
\end{figure}

From the perspective of MRB, the segmentation operation over set $S_{l}$ is equivalent to cutting the MRB positions into $Q$ segments and generating TEPs accordingly in the phase-$l$ reprocessing, as shown in Fig \ref{MRBsegmentation}. The $i$-th MRB segment is called sub-MRB $\mathbf{b}_{i}^{sub}$, defined as positions from boundary $\beta_{i}$ to $K$, i.e.
\begin{equation}
\mathbf{b}_{i}^{sub} = \left[\beta_{i},\beta_{i}+1,\ldots,K \right]
\end{equation}
for $0<i \leq Q$. Particularly $\mathbf{b}_{Q}^{sub}$ is exactly the MRB, defined as positions from $1$ to $K$. Let $S(\mathbf{e}_{i}^{l})$ denote the set of weight-$l$ TEPs which only have nonzero elements over positions $\mathbf{b}_{i}^{sub}$, thus the TEPs segment $S_{l{i}}$ can be easily obtained by
\begin{equation} \label{Segmentsgeneration}
S_{l_{i}} = S(\mathbf{e}_{i}^{l}) \backslash S(\mathbf{e}_{i-1}^{l})
\end{equation}
for $1<i\leq Q$. Particularly when $i=1$, the TEP segment $S_{l_{1}}$ is identical to $S(\mathbf{e}_{1}^{l})$.

The choice of boundaries during each reprocessing can greatly affect the trade-off between performance and complexity when some segments are discarded. We determine the boundaries by considering a deviation of the mean of reliability values over MRB. The mean of reliabilities serves as a benchmark for the boundary calculation, and the deviation enables boundaries to be changed adaptively according to the decoding process. At the beginning of each reprocessing we first estimate the reliability of the first boundary position ${\widetilde \alpha}_{\beta_{1}}$ as
\begin{equation} \label{1boundary}
{\widetilde \alpha}_{\beta_{1}} = \frac{1}{\lambda}E_{[1,K]}\cdot f(\bm{\widetilde \alpha},\mathcal{D}_{min})\text{,}
\end{equation}
where $E_{[a,b]}$ is the mean of ordered reliabilities $\bm{\widetilde \alpha}$ over positions from $a$ to $b$, $\mathcal{D}_{min}$ is the minimum WHD from checked codeword candidates to received sequence, and $\lambda$ is a given parameter. Boundary reliability ${\widetilde \alpha}_{\beta_{1}}$ is tightened and updated adaptively at each reprocessing phase according to the offset function  $f(\bm{\widetilde \alpha},\mathcal{D}_{min})$. The choice of $f(\bm{\widetilde \alpha},\mathcal{D}_{min})$ will be discussed in Section \ref{Simulation}.  

The first boundary $\beta_{1}$ is determined by finding the position over MRB whose reliability is closest to ${\widetilde \alpha}_{\beta_{1}}$. Then boundary $\beta_{i}$, $1<i<Q$, is sequentially determined as the position over $[1,2,\ldots,\beta_{i-1}-1]$ whose reliability is closest to ${\widetilde \alpha}_{\beta_{i}}$ which is estimated as
\begin{equation} \label{otherboundary}
{\widetilde \alpha}_{\beta_{i}} = \frac{1}{\lambda}E_{[1,\beta_{i-1}-1]}\cdot f(\bm{\widetilde \alpha},\mathcal{D}_{min})\text{.}
\end{equation}

The value of $\lambda$ affects the positions of all $Q$ boundaries, Furthermore with discarding rule, the trade-off between complexity and performance can be adjusted by choosing different $\lambda$ value.

\subsection{Discarding and stopping rules} \label{discarding&stopping}
In order-$m$ OSD decoding, all the weight-$l$ TEPs are checked in phase-$l$ $(0\leq l\leq m)$ reprocessing. Thus some search strategies can be used to improve the checking efficiency. It is proved that for a reliability-ordered hard-decision sequence $\mathbf{\widetilde{y}}$, the following inequalities holds \cite{Fossorier1995OSD},
\begin{equation} 
\textup{P}_{\textup{e}}\left(i;N\right)<\textup{P}_{\textup{e}}\left(i+1;N\right)
\end{equation}
and
\begin{equation} \label{Probability}
\textup{P}_{\textup{e}}\left(h,i;N\right) \leq \textup{P}_{\textup{e}}\left(i,j;N\right) \leq \textup{P}_{\textup{e}}\left(i;N\right)
\end{equation}
for $1<h<i<j<N$, where $N$ is the sequence length and $\textup{P}_{\textup{e}}\left(i;N\right)$ is the probability that the hard-decision of $i$-th symbol is in error. Equivalent results hold for any number of positions considered. Therefore, one of the regular search orders is to start checking TEPs from the least reliable positions with least weight\cite{valembois2002comparison}.

From (\ref{boundary position}) and (\ref{Probability}) it can be concluded that the first TEPs segments have the highest checking priority within the same phase reprocessing, and those of the rest are diminishing.
Therefore, a promising scheme for order-$m$ decoding is to conduct reprocessing $(m+1)$ times in an ascending phase order from 0 to $m$ and checking TEPs segments individually from $S_{l_{1}}$ to $S_{l_{Q}}$ in phase-$l$ $(1\leq l\leq m)$ reprocessing. Since the last few segments in every reprocessing procedures have the TEPs with least opportunities for successful re-encoding, some of them can be discarded to reduce the decoding complexity.

We introduce a segments discarding rule utilizing local lower bounds $\mathcal{D}^{lower}_{l}$ of WHD in each phase-$l$ reprocessing ($1\leq l\leq m$). When current minimum WHD $\mathcal{D}_{min}$ is lower than the local lower bound, all the remaining unprocessed segments in corresponding reprocessing phase will be discarded. Thus the segments are discarded if the following condition is satisfied
\begin{equation}
\mathcal{D}_{min}<\mathcal{D}^{lower}_{l} \text{.}
\end{equation}
$\mathcal{D}^{lower}_{l}$ is estimated from the first checked TEP $\mathbf{e}'$ in segments $S_{l_{i}}$, and it is tightened and updated in every TEP segments checking procedure. When the reprocessing starts checking TEPs from a new segments, the $\mathcal{D}^{lower}_{l}$ is updated by
\begin{equation}
\mathcal{D}^{lower}_{l} = \mathcal{L}\left(1+\tau \sigma(\bm{\widetilde \alpha}) \frac{E_{[K+1,N]}}{E_{[1,K]}} \right)\text{,}
\end{equation}
where
\begin{equation} \label{sumReliability}
\mathcal{L} = \sum_{\substack{0<i<K \\ e'_{i}\neq 0}} \widetilde{\alpha}_{i}
\end{equation}
is the sum of reliabilities over nonzero positions of $\mathbf{e}'$, $\sigma(\bm{\widetilde \alpha})$ is the standard deviation of ordered reliabilities $\bm{\widetilde \alpha}$, and $\tau$ is a parameter that can adjust the trade-off between performance and complexity. In \cite{Wu2007OSDMRB}, a similar approach was used to estimate TEP likelihood from WHD $\mathcal{D}_{min}$.

\begin{figure}
	\begin{center}
		\includegraphics[scale=0.28] {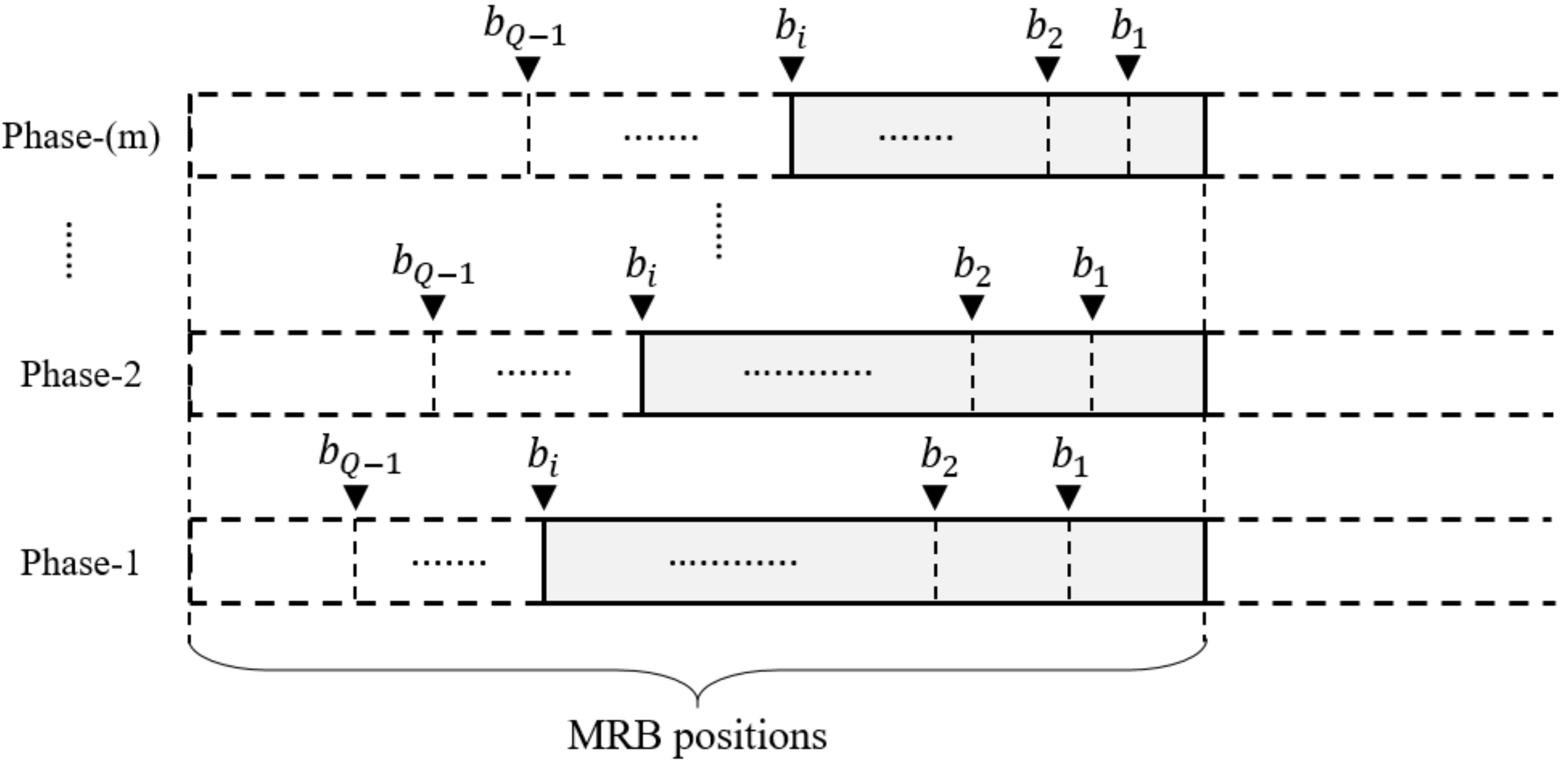}
		\caption{Decoding Scheme. Light colored blocks represent the segments that was discarded, while dark-colored ones are retained.}
		\label{Scheme}
	\end{center}
\end{figure}

Assuming that segments $\{S_{l_Q},S_{l_{Q-1}},\ldots,S_{l_{i}} \}$ are discarded in phase-$l$ reprocessing, the combining segmentation and discarding scheme is depicted in Fig. \ref{Scheme}, where discarded segments are indicated by light colored blocks.

Another stopping rule utilizing the first boundary $\beta_{1}$ is used to terminate the decoding in advance. During phase-$l$ reprocessing, the decoding stops and outputs the result immediately if the following condition is satisfied 
\begin{equation}
\beta_{1} \geq K - l + 1
\end{equation}
for $0<l\leq m$. This is because that if sub-MRB $\mathbf{b}_{1}^{sub}$ does not have enough positions to generate a weight-$l$ TEP, the decoding codeword is close to the ideal output and no further decoding needs to be conducted.

We present the complete decoding algorithm combining segmentation-discarding and stopping rules in Algorithm \ref{algorithmCode}.

\begin{algorithm}[t]
	\caption{Proposed SDD Algorithm}
	\label{algorithmCode}
	\begin{algorithmic} [1]
		\REQUIRE ~~\\
		Generator matrix $\mathbf{G}$, received sequence $\mathbf{r}$\\
		Order $m$, segments number $Q$,
		and parameters $\lambda$ and $\tau$
		\ENSURE ~~\\
		 Optimal codeword estimation $\hat{\mathbf{c}}_{opt}$
		
		~~\\
		
		\STATE Calculate reliability value $\alpha_{i} = |r_{i}|$
		\STATE First permutation: 
     		$\bm {\alpha}'= \pi_{1}\left(\bm{\alpha}\right)$, $\mathbf {r}'=\pi_{1}\left(\mathbf{r}\right)$, $\mathbf {G}'=\pi_{1}\left(\mathbf{G}\right)$
		\STATE Gaussian elimination and second permutation:
			$\mathbf{\widetilde a} = \pi_{2}(\mathbf {a}')$, $\mathbf{\widetilde r} = \pi_{2}(\mathbf {r}')$, $\mathbf{\widetilde G} = \pi_{2}(\mathbf {G}')$
		\STATE Perform hard-decision: 			
			$\widetilde{y}_{i}=
			\begin{cases}
				1& \text{for} \ \widetilde{r}_{i}<0\\
				0& \text{for} \ \widetilde{r}_{i}\geq 0
			\end{cases}\ $
		\STATE $//$\textbf{Phase-0 reprocessing}
		\STATE Calculate  $\mathbf{\widetilde{c}}_{opt} = \mathbf{\widetilde{y}}_{B}\mathbf{\widetilde{G}} $ and $\mathcal{D}_{min} = \sum\limits_{\substack{0<i<N \\ \widetilde{c}_{opt,i}\neq \widetilde{y}_{i}}} \widetilde{a}_{i}$
		\STATE $//$\textbf{Phase-$l$ reprocessing with $Q$ segments}
		\FOR{$l=1:m$}
		\FOR{$i=1:Q$}
		\STATE Determine the boundary $\beta_{i}$ through \\ ${\widetilde \alpha}_{\beta_{i}} = \frac{1}{\lambda}E_{[1,\beta_{i-1}-1]}\cdot f(\bm{\widetilde \alpha},\mathcal{D}_{min})$
		\IF{$\beta_{i}\geq K-l+1$}
		\RETURN $\hat{\mathbf{c}}_{opt} = \pi_{1}^{-1}(\pi_{2}^{-1}(\mathbf{\widetilde c}_{opt}))$
		\ENDIF
		\STATE Generate TEP segment $S_{l_{i}}$
		\STATE Calculate $\mathcal{L}$ and $\mathcal{D}^{lower}_{l} = \mathcal{L}\left(1+\tau \sigma(\bm{\widetilde \alpha})\frac{E_{[K+1,N]}}{E_{[1,K]}}\right)$
		\IF{$\mathcal{D}_{min}<\mathcal{D}^{lower}_{l}$}
		\STATE \textbf{break}
		\ENDIF
		\STATE Check all TEPs $\mathbf{e}$ from $S_{l_{i}}$ by re-encoding $\mathbf{\widetilde c}_{e} = \left(\mathbf{\widetilde y}_{B}\oplus \mathbf{e}\right)\mathbf{\widetilde G}$ and evaluating $\mathcal{D}_{e}$.
		Find local optimum estimation $\mathbf{\widetilde c}^{local}_{opt}$ with distance $\mathcal{D}^{local}_{min}$ for $S_{l_{i}}$
		\IF{$\mathcal{D}^{local}_{min}<\mathcal{D}_{min}$}
		\STATE $\mathcal{D}_{min}=\mathcal{D}^{local}_{min}$, $\mathbf{\widetilde c}_{opt}=\mathbf{\widetilde c}^{local}_{opt}$

		\ENDIF
		\ENDFOR		
		\ENDFOR 
		\RETURN $\hat{\mathbf{c}}_{opt} = \pi_{1}^{-1}(\pi_{2}^{-1}(\mathbf{\widetilde c}_{opt}))$
	\end{algorithmic}
\end{algorithm}

\section{Computational Complexity} \label{Computational Complexity}
We estimate the algorithm complexity by evaluating the number of floating point operations (FLOPs) and binary operations (BOPs) of each step. The total computational complexity is mainly dependent on following terms:
\begin{itemize}
	
	\item Sorting (the first permutation): Merge sort algorithm can efficiently generate and perform the first permutation with average complexity of $O(N\log N)$ FLOPs\cite{Fossorier1995OSD}.
	
	\item Gaussian elimination: The operation to obtain systematical generation matrix $\mathbf{\widetilde{G}}$ from $\mathbf{G}'$ can be done with $O(N\min(K,N-K)^2)$\cite{Fossorier1995OSD}. 
	
	\item Re-encoding: Re-encoding $\mathbf{\widetilde{c}}_{\mathbf{e}} = (\mathbf{\widetilde y}_{B}\oplus \mathbf{e})\mathbf{\widetilde G}$ uses $K$ sign operations and $N-K$ parallel $K$ XOR operations \cite{Fossorier1995OSD}, which can be represented as $O(K+K(N-K))$ BOPs.
	
	\item Number of candidates: For OSD-based decoding, the total number of checked candidates $N_{a}$ greatly affect the complexity since $N_{a}$ times of re-encoding is required.
	
	\item Segment boundaries and distance lower bound: The searching of $\beta_{i}$ can be regarded as one-dimension look-up table operation with $O(K)$, thus the total cost of $Q$ boundaries calculation is $O(KQ)$ FLOPs. While distance lower bound is simply calculated with complexity of $O(1)$ FLOP in every segments.
\end{itemize}

In a complete decoding procedure, the sorting and Gaussian elimination is performed once, the re-encoding repeats $N_{a}$ times in reprocessing, the boundaries are calculated $m$ times, and the distance lower bound is updated $mQ$ times. Therefore, the total computational complexity can be estimated as
\begin{equation}
\begin{split}
C_{total} \approx & N\log N + N\min(N,N-K)\\
& + N_{a}(K+K(N-K)) + (K+1)mQ.
\end{split}
\end{equation}
The last term, e.i., $(K+1)mQ$ can be ignored since it is too small in comparison with the other terms when Q is not large. This implies that the complexity due to the segmentation and discarding is negligible. 

As a comparison, we also derive the extra computation of the Fast OSD algorithm from \cite{NewOSD-5GNR}. The probabilistic necessary condition (PNC) in \cite{NewOSD-5GNR} requires $(N-K)$ parallel $N$ XOR operations and is checked at least once during each reprocessing. The probabilistic sufficient condition (PSC) requires approximately one FLOP and is checked for each TEP \cite{NewOSD-5GNR}. Therefore, the overall extra computation due to PNC and PSC in the Fast OSD \cite{NewOSD-5GNR} is at least
\begin{equation} \label{extraFast}
mN(N-K) + N_{a} \text{.}
\end{equation}
Compared with (\ref{extraFast}), our proposed algorithm is more efficient in terms of the computational complexity since only $(K+1)mQ$ extra operations are introduced.

\section{Simulation Results and Comparisons} \label{Simulation}
In this section, we present several simulation results and comparisons for length-128 extended BCH (eBCH) codes with different rates. The form of offset function $f(\bm{\widetilde \alpha},\mathcal{D}_{min})$ in (\ref{1boundary}) will significantly affect the performance of the proposed algorithm. By simulation, we find that the best decoding efficiency is obtained when the offset function has the following form
\begin{equation} \label{Offsetfunction}
f(\bm{\widetilde \alpha},\mathcal{D}_{min}) = \frac{\mathcal{D}_{min}}{E_{[1,N]}}\text{.}
\end{equation}
Substituting the offset function (\ref{Offsetfunction}) into (\ref{1boundary}), the first boundary is the position whose reliabilities is closest to 
\begin{equation} 
{\widetilde \alpha}_{\beta_{i}} = E_{[1,\beta_{i-1}]}\frac{\mathcal{D}_{min}}{\lambda E_{[1,N]}}\text{.}
\end{equation}

\begin{figure}
	\begin{center}
		\includegraphics[scale=0.5] {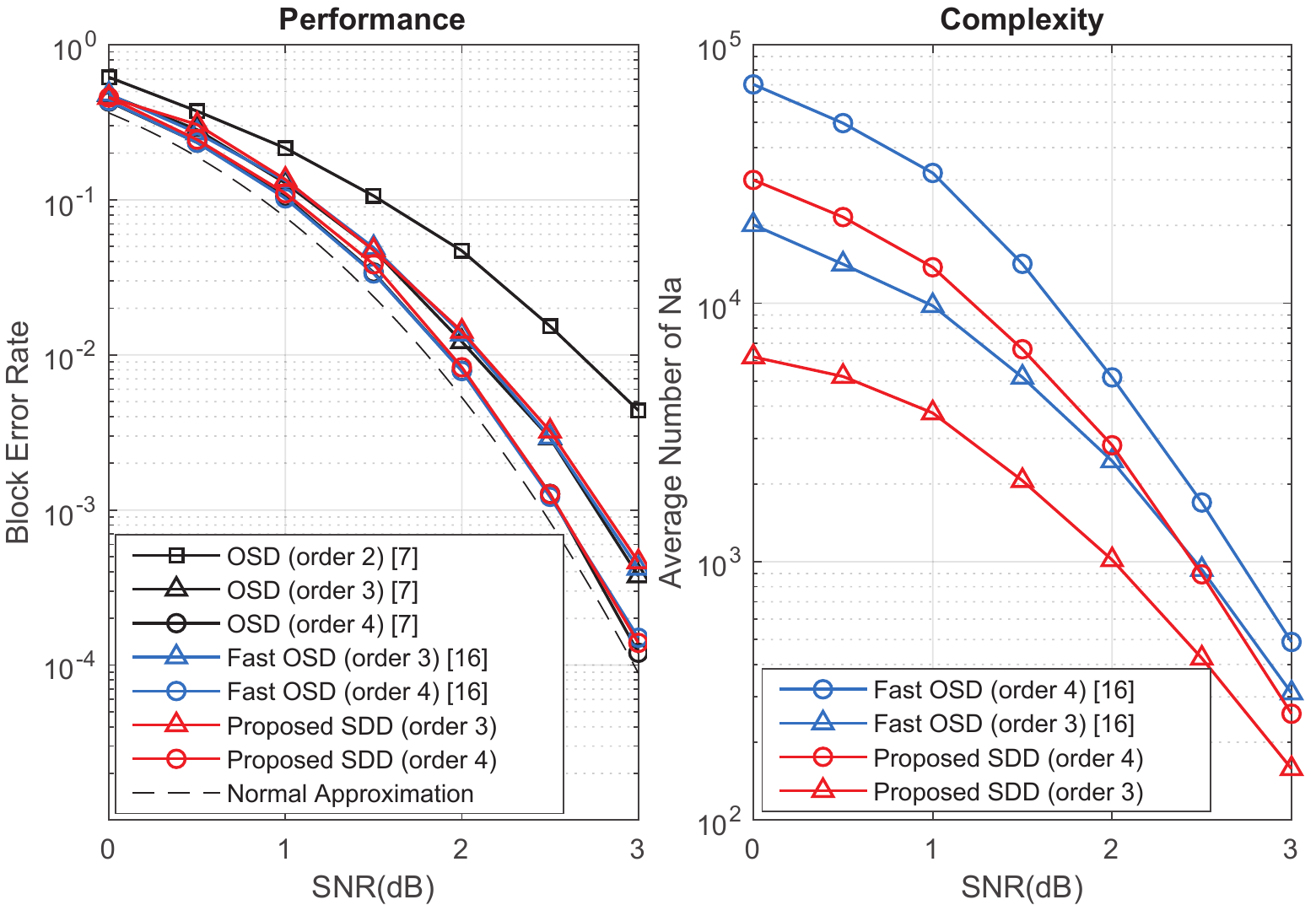}
		\caption{Performance and complexity comparison in decoding the (128,64,22) eBCH code.}
		\label{64}
	\end{center}
\end{figure}

The performance and complexity of various decoders for $(128,64,22)$ eBCH code is depicted in Fig. \ref{64}. For our proposed algorithm, we set segment number $Q=22$, parameters $\lambda = 10.5$, and $\tau = 7$ in order-4 decoding and $\tau = 9.25$ in order-3 decoding. The original OSD algorithm \cite{Fossorier1995OSD}, the recent proposed OSD fast approach \cite{NewOSD-5GNR}, and the normal approximation of the Polyanskiy\text{-}Poor\text{-}Verd\'{u} (PPV) \cite{PPV2010l} are included in simulation as benchmarks for comparison. From simulation results, it can be seen that our proposed SDD algorithm exhibits a nearly identical performance compared to other simulated counterparts. However, the complexity in terms of the average number of checked candidates is different. Compared to fast OSD approach, our algorithm requires less than half of the TEP candidates.

\begin{figure}
	\begin{center}
		\includegraphics[scale=0.5] {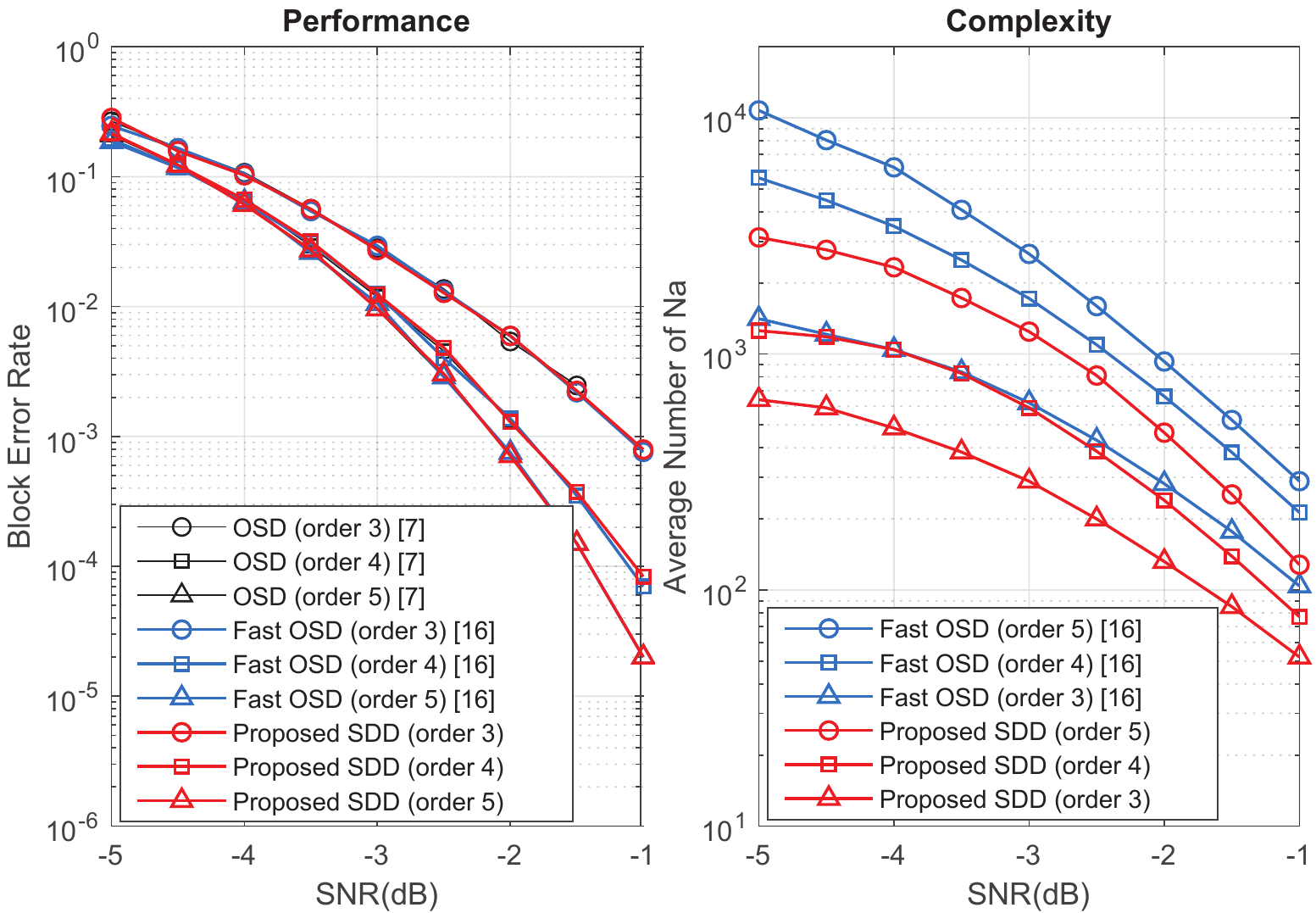}
		\caption{Performance and complexity comparison in decoding the (128,22,48) eBCH code.}
		\label{22}
	\end{center}
\end{figure}

Same simulation is conducted for a lower coding rate with $(128,22,48)$ eBCH code. For this case, we set $Q=16$, parameters $\lambda = 23$ and $\tau = 7.25,9,11.25$ for oder-5, order-4 and order-3 decoding, respectively. As shown in Fig. \ref{22}, the performance and complexity is compared for different approaches and significant improvement is brought by the proposed SDD algorithm. At low SNRs, our decoding achieves the same performance using three times less the number of candidates than fast OSD, and particularly significant complexity reduction can be observed at high SNRs as well. Note that the PPV normal approximation is not included in simulation because of its inaccuracy at low coding rate.

\begin{figure}
	\begin{center}
		\includegraphics[scale=0.492] {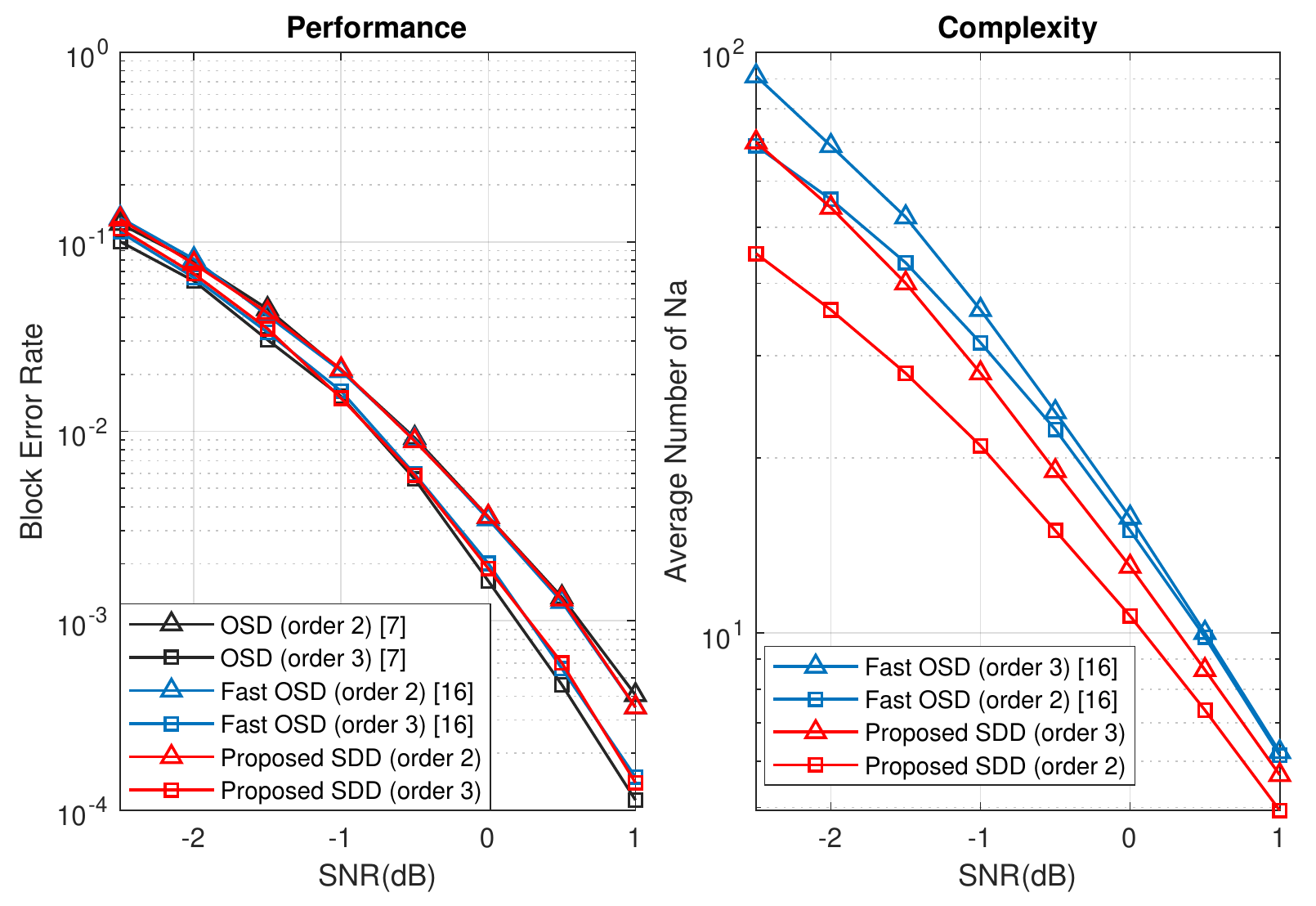}
		\caption{Performance and complexity comparison in decoding the (64,16,24) eBCH code.}
		\label{16}
	\end{center}
\end{figure}

The simulation results of decoding (64,16,24) eBCH code is depicted in Fig. \ref{16}. For the proposed SDD, we set $Q=16$,$\lambda=13$, $\tau=5$ and $\tau=5.5$ for order-3 and order-2 decoding, respectively. In this 64-length regime, the proposed SDD also outperforms the fast OSD in terms of the decoding complexity, with the near-optimal performance achieved. The average numbers of $N_{a}$ in decoding the above three codes are recorded in Table \ref{tab:Na64}, Table \ref{tab:Na22} and Table \ref{tab:Na16}.

\begin{figure}
	\begin{center}
		\includegraphics[scale=0.5] {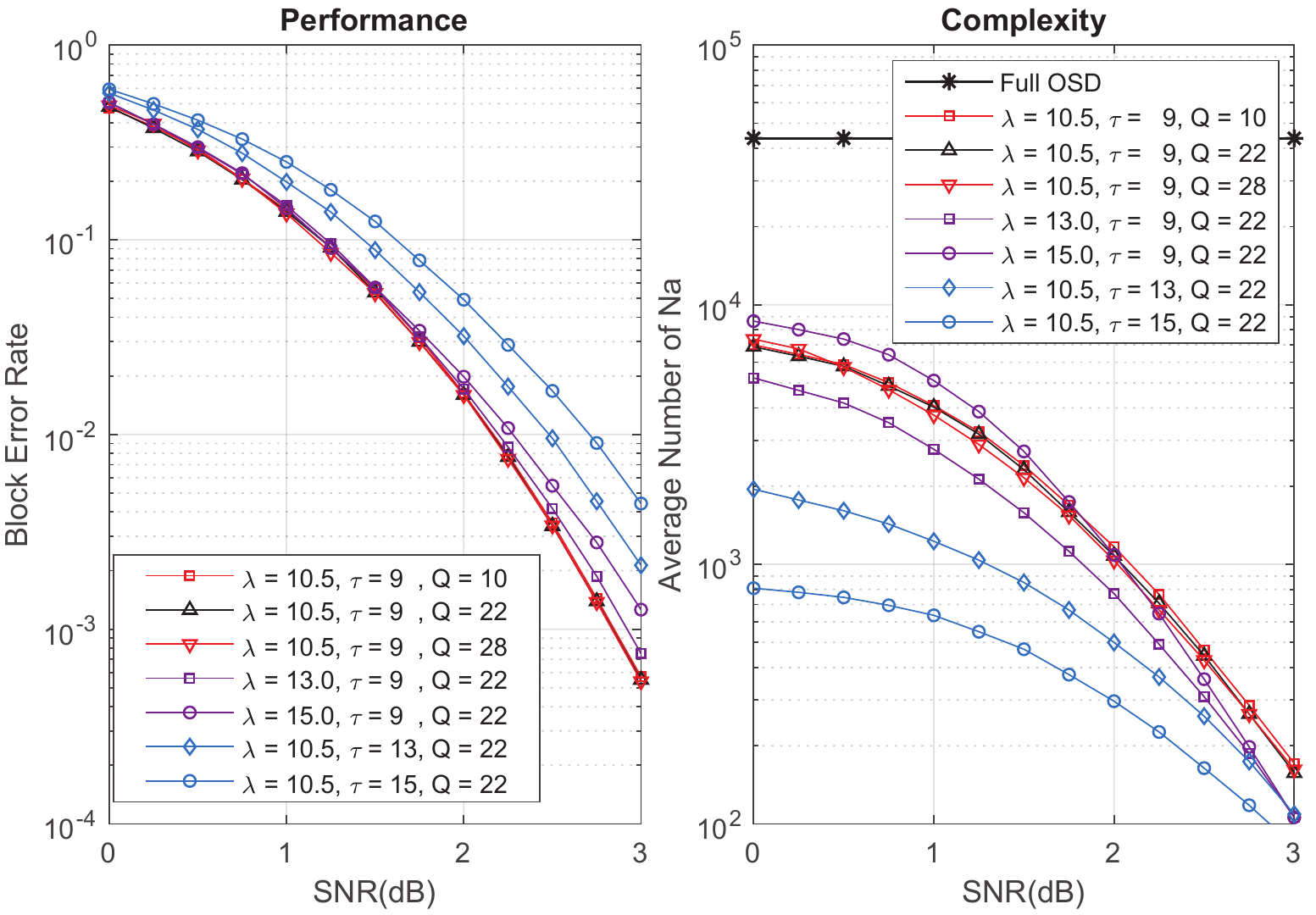}
		\caption{Order-3 decoding of the (128,64,22) eBCH code with different parameter $\lambda$, $\tau$ and segments number $Q$.}
		\label{Trad}
	\end{center}
\end{figure}

For completeness, we have also conducted a study of the impact of different segments number $Q$, parameter $\lambda$ and $\tau$ values on performance and complexity in order-3 decoding for (128,64,22) eBCH code. As depicted in Fig \ref{Trad}, it can be seen that the performance decreases gradually with increasing $\tau$, and the average candidates number $N_{a}$ is reduced accordingly. $\lambda$ affects the performance at high SNRs and simulation advises that is there is an optimal value for $\lambda$.
Changing $Q$ also affects the decoding efficiency because more segments bring more discarding options. Choosing different parameters can adjust the trade-off between performance and complexity to meet the needs of different decoding requirements.

\begin{figure}
	\begin{center}
		\includegraphics[scale=0.45] {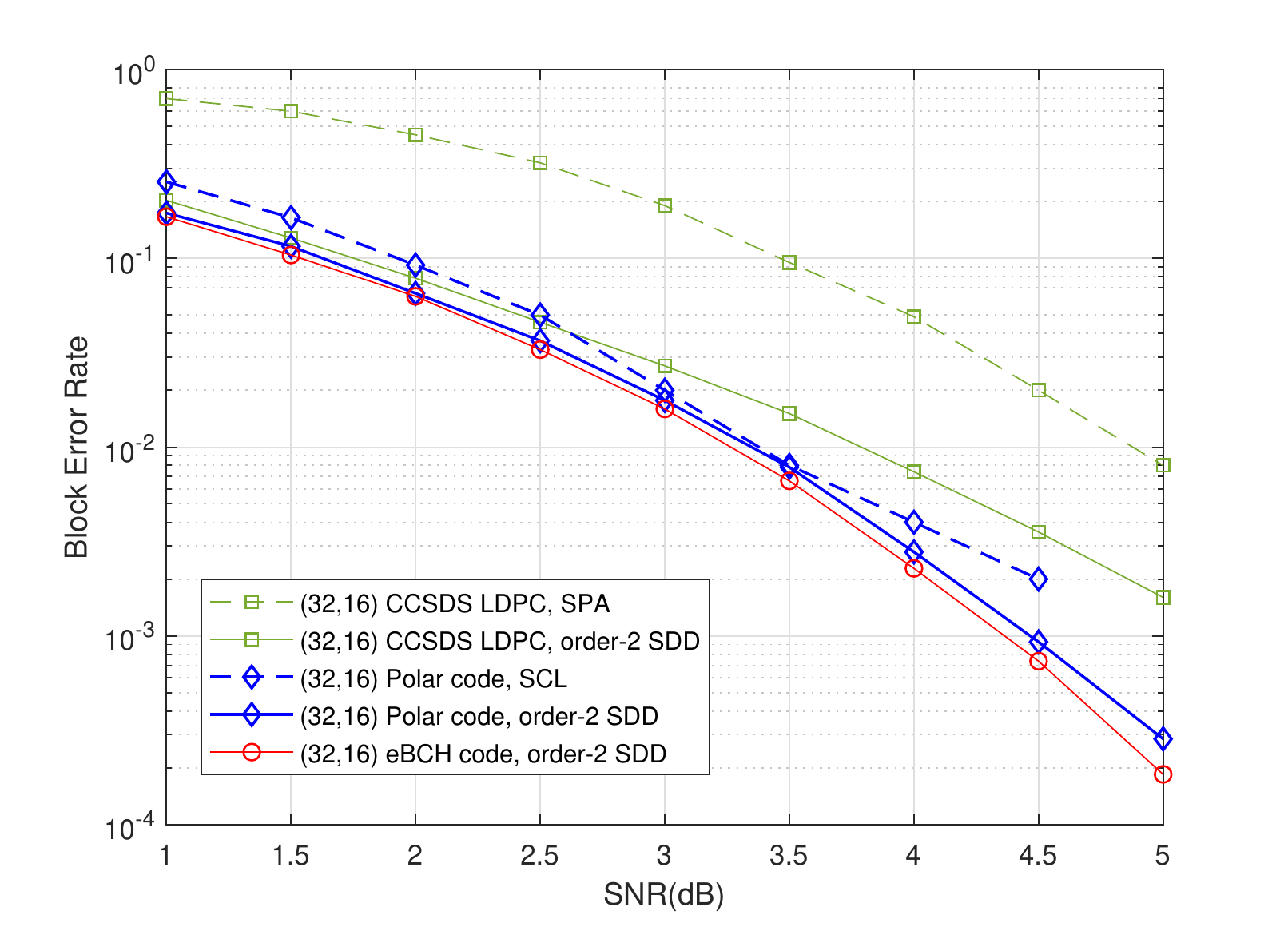}
		\caption{Decoding performance comparison for length-32 codes}
		\label{32}
	\end{center}
\end{figure}

Apart from the class of BCH codes, the proposed SDD also has the potential to be a universal decoding approach for all linear block codes in the short block-length regime. We compared the decoding performance for various length-32 codes with fixed coding rate 0.5, as depicted in Fig. \ref{32}. The (32,16) eBCH code, the CCSDS standard LDPC code, and the (32,16) Polar codes are decoded by the SDD decoder and also their corresponding traditional decoder (SPA for LDPC codes and SCL for Polar codes)\cite{channelcodes}. From the simulation results, BCH code performs best among these three codes, and the Polar code is slightly inferior. The block-error-rate performance of LDPC code is the worst, only reaches $10^{-3}$ using SDD and reaches $10^{-2}$ using SPA. For decoding both the LDPC code and Polar code, the proposed SDD outperforms their traditional decoder.

\begin{table} [t]
	\centering
	\caption{Comparison of average $N_{a}$ between fast OSD and proposed algorithm for (128,64,22) eBCH code}
	\label{tab:Na64}
	\begin{tabular}{|c|c|c|c|c|}
		\hline
		SNR(dB)& 0 & 1 & 2 & 3\\
		\hline
		\hline
		Order-3 fast OSD  & 20107 & 9775 & 2452 & 310\\
		Order-3 proposed algorithm & 6194 & 3762 & 1016 & 158\\
		\hline
		\hline
		Order-4 fast OSD  & 70262 & 31917 & 5164 & 489\\
		Order-4 proposed algorithm  & 29992 & 13777 & 2821 & 258\\
		\hline
		
	\end{tabular}
\end{table}

\begin{table} [t]
	\centering
	\caption{Comparison of average $N_{a}$ between fast OSD and proposed algorithm for (128,22,48) eBCH code}
	\label{tab:Na22}
	\begin{tabular}{|c|c|c|c|c|c|}
		\hline
		SNR(dB)& -5 & -4 & -3 & -2 & -1\\
		\hline
		\hline
		Order-3 fast OSD & 1409 & 1043 & 621 & 282 & 104 \\
		Order-3 proposed algorithm & 640 & 485 & 289 & 132 & 52\\
		\hline
		\hline
		Order-4 fast OSD & 5567 & 3475 & 1718 & 661 & 312\\
		Order-4 proposed algorithm & 1255 & 1072 & 591 & 240 & 77\\
		\hline
		\hline		
		Order-5 fast OSD& 10753 & 6172 & 2635 & 929 & 289\\
		Order-5 proposed algorithm & 3116 & 2328 & 1243 & 464 & 128\\
		\hline
		
	\end{tabular}
\end{table}

\begin{table} [t]
	\centering
	\caption{Comparison of average $N_{a}$ between fast OSD and proposed algorithm for (64,16,24) eBCH code}
	\label{tab:Na16}
	\begin{tabular}{|c|c|c|c|c|}
		\hline
		SNR(dB)& -2 & -1 & 0 & 1\\
		\hline
		\hline
		Order-2 fast OSD  & 55.9 & 31.6 & 15.1 & 6.1\\
		Order-2 proposed algorithm  & 36.4 & 21.0 & 10.7 & 4.9\\
		\hline
		\hline
		Order-3 fast OSD  & 69.8 & 36.0 & 15.9 & 6.2\\
		Order-3 proposed algorithm  & 54.4 & 28.2 & 13.0 & 5.7\\
		\hline
		
	\end{tabular}
\end{table}

\section{Conclusion} \label{Conclusion}
In this paper we proposed a new fast segmentation-discarding decoding (SDD) algorithm for short block-length codes based on ordered reliability. Two techniques were combined in the proposed approach: 1) an adaptive segmentation and discarding rule to discard unpromising TEPs, and 2) a stopping rule to terminate the decoding when good estimation has been found.

From simulation results, we conclude that the proposed algorithm can significantly reduce the decoding complexity of OSD for multiple rate short block-length eBCH codes. By adjusting parameters, the trade-off between performance and complexity can be obtained. In addition, the proposed algorithm has the potential to be a universal decoding approach for any linear codes in the short block-length regime with near-optimal performance guaranteed.


%\section*{Acknowledgment}



\bibliographystyle{IEEEtran}
\bibliography{IEEEabrv,MyCollection}

\end{document}